\documentclass[12pt]{article}
\usepackage{euler}
\usepackage{amsmath}
\usepackage{amssymb}
\usepackage{amstext}
\usepackage{amscd}
\usepackage{amsfonts}
\DeclareMathAlphabet{\EuFrak}{U}{euf}{m}{n}
\DeclareMathAlphabet{\EuScript}{U}{eus}{m}{n}

\title{{\bf World Sheet Superstring and Superstring Field Theory: a new
solution using Ultradistributions of Exponential Type}
\thanks{\it{This work was partially supported by Consejo
Nacional
de Investigaciones Cient\'{\i}ficas and Comisi\'{o}n de
Investigaciones Cient\'{\i}ficas de la Pcia. de Buenos
Aires;
Argentina.${\dag}$Deceased}}}
\author{C.G.Bollini$^{\dag}$, A. L. De Paoli, M.C.Rocca\\
Departamento de F\'{\i}sica, Fac. de Ciencias Exactas,\\
Universidad Nacional de La Plata.\\
C.C. 67 (1900) La Plata. Argentina.}

\date{October 24 , 2008}

\begin{document}

\maketitle

\vspace{-5mm}

\begin{abstract}

In this paper we show that 
Ultradistributions of Exponential Type
(UET) are appropriate for the description
in a consistent way world sheet superstring and 
superstring field theories.
A new Lagrangian for the closed world sheet 
superstring is obtained.
We also show that the superstring field 
is a linear superposition of UET of compact support (CUET),
and give the notion of anti-superstring.
We evaluate the propagator  for the string field,
and calculate the convolution of two of them.


\end{abstract}

\newpage

\renewcommand{\theequation}{\arabic{section}.\arabic{equation}}

\section{Introduction}

In a serie of papers \cite{tp1,tp2,tp3,tp4,tp5}
we have shown that Ultradistribution theory of 
Sebastiao e Silva  \cite{tp6,tp7,tp8} permits a significant advance in the treatment 
of quantum field theory. In particular, with the use of the 
convolution of Ultradistributions we have shown that it  is possible
to define a general product of distributions ( a product in a ring
with divisors of zero) that sheds new ligth on  the question of the divergences
in Quantum Field Theory. Furthermore, Ultradistributions of Exponential Type  
are  adequates to describe
Gamow States and exponentially increasing fields in Quantum 
Field Theory \cite{tp9,tp10,tp11}.

Ultradistributions also have the
advantage of being representable by means of analytic functions.
So that, in general, they are easier to work with  and,
as we shall see, have interesting properties. One of these properties
is that Schwartz's tempered distributions are canonical and continuously
injected into  Ultradistributions of Exponential Type
and as a consequence the Rigged
Hilbert Space  with tempered distributions is  canonical and continuously
included
in the Rigged Hilbert Space with  Ultradistributions of 
Exponential Type.

Another interesting property is that the space of UET is 
reflexive under the operation of Fourier transform (in a similar way
to that tempered distributions of Schwartz)

In two recent papers (\cite{tq1,ts2})  we have shown
that Ultradistributions of Exponential type
provide an adecuate framework 
for a consistent treatment of 
string and string field theories. In particular, a general
state of the closed string is 
represented by UET of compact support,
and as a consequence the string field is a linear combination
of UET of compact support.

In this paper we extend the treatment to 
world sheet superstrings.

This paper is organized as follows:
in sections 2 and 3 we define the Ultradistributions of Exponential Type
and their Fourier transform. They
are  part of a Guelfand's Triplet ( or Rigged Hilbert Space \cite{tp12} )
together with their respective duals and a ``middle term'' Hilbert
space. 
In sections 4 and 5  we give the main results of ref.\cite{tq1}
for the bosonic string used in the present work.
In section 6 we obtain  a  expression for the
Lagrangian of a closed world sheet superstring.
We give a solution of the equations of motion
and show that
the Lagrangian is supersymmetrically invariant
In section 7 we give a new representation for the states 
of the string using CUET.
In section 8 we give  expressions for the field of the superstring,
the superstring field propagator, the creation and anihilation
operators of a superstring and a anti-superstring.
In section 9, we give expressions for the non-local action of a free string
and a non-local interaction lagrangian for the string 
field in a way similar to that given  
in Quantum Field Theory.
Also we show how to evaluate the convolution
of two superstring field propagators.
Finally, section 10 is reserved for a discussion of the principal results.

\section{Ultradistributions of Exponential Type}

Let ${\cal S}$ be the Schwartz space of rapidly decreasing test functions. 
Let ${\Lambda}_j$ be the region of the complex plane defined as:
\begin{equation}
\label{er2.1}
{\Lambda}_j=\left\{z\in\boldsymbol{\mathbb{C}} :
|\Im(z)|< j : j\in\boldsymbol{\mathbb{N}}\right\}
\end{equation}
According to ref.\cite{tp6,tp8} the space of test functions $\hat{\phi}\in
{\large{V}}_j$ is
constituted by all entire analytic functions of ${\cal S}$ for which
\begin{equation}
\label{er2.2}
||\hat{\phi} ||_j=\max_{k\leq j}\left\{\sup_{z\in{\Lambda}_j}\left[e^{(j|\Re (z)|)}
|{\hat{\phi}}^{(k)}(z)|\right]\right\}
\end{equation}
is finite.\\
The space ${\large{Z}}$ is then defined as:
\begin{equation}
\label{er2.3}
{\large{Z}} =\bigcap_{j=0}^{\infty} {\large{V}}_j
\end{equation}
It is a complete countably normed space with the topology generated by
the system of semi-norms $\{||\cdot ||_j\}_{j\in \mathbb{N}}$.
The dual of ${\large{Z}}$, denoted by
${\large{B}}$, is by definition the space of ultradistributions of exponential
type (ref.\cite{tp6,tp8}).
Let ${S}$ be the space of rapidly decreasing sequences. According to
ref.\cite{tp12} ${S}$ is a nuclear space. We consider now the space of
sequences ${P}$ generated by the Taylor development of
$\hat{\phi}\in{\large{Z}}$
\begin{equation}
\label{er2.4}
{P}=\left\{{Q} : {Q}
\left(\hat{\phi}(0),{\hat{\phi}}^{'}(0),\frac {{\hat{\phi}}^{''}(0)} {2},...,
\frac {{\hat{\phi}}^{(n)}(0)} {n!},...\right) : \hat{\phi}\in{Z}\right\}
\end{equation}
The norms that define the topology of ${P}$ are given by:
\begin{equation}
\label{er2.5}
||\hat{\phi} ||^{'}_p=\sup_n \frac {n^p} {n} |{\hat{\phi}}^n(0)|
\end{equation}
${P}$ is a subspace of ${S}$ and therefore is a nuclear space.
As the norms $||\cdot ||_j$ and $||\cdot ||^{'}_p$ are equivalent, the correspondence
\begin{equation}
\label{er2.6}
{\large{Z}}\Longleftrightarrow {P}
\end{equation}
is an isomorphism and therefore ${Z}$ is a countably normed nuclear space.
We can define now the set of scalar products
\[<\hat{\phi}(z),\hat{\psi}(z)>_n=\sum\limits_{q=0}^n\int\limits_{-\infty}^{\infty}e^{2n|z|}
\overline{{\hat{\phi}}^{(q)}}(z){\hat{\psi}}^{(q)}(z)\;dz=\]
\begin{equation}
\label{er2.7}
\sum\limits_{q=0}^n\int\limits_{-\infty}^{\infty}e^{2n|x|}
\overline{{\hat{\phi}}^{(q)}}(x){\hat{\psi}}^{(q)}(x)\;dx
\end{equation}
This scalar product induces the norm
\begin{equation}
\label{er2.8}
||\hat{\phi}||_n^{''}=[<\hat{\phi}(x),\hat{\phi}(x)>_n]^{\frac {1} {2}}
\end{equation}
The norms $||\cdot ||_j$ and $||\cdot ||^{''}_n$ are equivalent, and therefore
${\large{Z}}$ is a countably hilbertian nuclear space.
Thus, if we call now ${{\large{Z}}}_p$ the completion of
${\large{Z}}$ by the norm $p$ given in (\ref{er2.8}), we have:
\begin{equation}
\label{er2.9}
{\large{Z}}=\bigcap_{p=0}^{\infty}{{\large{Z}}}_p
\end{equation}
where
\begin{equation}
\label{er2.10}
{{\large{Z}}}_0=\boldsymbol{H}
\end{equation}
is the Hilbert space of square integrable functions.\\
As a consequence the ``nested space''
\begin{equation}
\label{er2.11}
{\Large{U}}=\boldsymbol{(}{\large{Z}},
\boldsymbol{H}, {\large{B}}\boldsymbol{)}
\end{equation}
is a Guelfand's triplet (or a Rigged Hilbert space=RHS. See ref.\cite{tp12}).

Any Guelfand's triplet
${\Large{G}}=\boldsymbol{(}\boldsymbol{\Phi},
\boldsymbol{H},\boldsymbol{{\Phi}^{'}}\boldsymbol{)}$
has the fundamental property that a linear and symmetric operator
on $\boldsymbol{\Phi}$, admitting an extension to a self-adjoint
operator in
$\boldsymbol{H}$, has a complete set of generalized eigen-functions
in $\boldsymbol{{\Phi}^{'}}$ with real eigenvalues.

${\large{B}}$ can also be characterized in the following way
( refs.\cite{tp6},\cite{tp8} ): let ${{E}}_{\omega}$ be the space of
all functions $\hat{F}(z)$ such that:

${\Large {\boldsymbol{I}}}$-
$\hat{F}(z)$ is analytic for $\{z\in \boldsymbol{\mathbb{C}} :
|Im(z)|>p\}$.

${\Large {\boldsymbol{II}}}$-
$\hat{F}(z)e^{-p|\Re(z)|}/z^p$ is bounded continuous  in
$\{z\in \boldsymbol{\mathbb{C}} :|Im(z)|\geqq p\}$,
where $p=0,1,2,...$ depends on $\hat{F}(z)$.

Let ${N}$ be:
${N}=\{\hat{F}(z)\in{{E}}_{\omega} :\hat{F}(z)\; \rm{is\; entire\; analytic}\}$.
Then ${\large{B}}$ is the quotient space:

${\Large {\boldsymbol{III}}}$-
${\large{B}}={{E}}_{\omega}/{N}$

Due to these properties it is possible to represent any ultradistribution
as ( ref.\cite{tp6,tp8} ):
\begin{equation}
\label{er2.12}
\hat{F}(\hat{\phi})=<\hat{F}(z), \hat{\phi}(z)>=\oint\limits_{\Gamma} \hat{F}(z) \hat{\phi}(z)\;dz
\end{equation}
where the path ${\Gamma}$ runs parallel to the real axis from
$-\infty$ to $\infty$ for $Im(z)>\zeta$, $\zeta>p$ and back from
$\infty$ to $-\infty$ for $Im(z)<-\zeta$, $-\zeta<-p$.
( $\Gamma$ surrounds all the singularities of $\hat{F}(z)$ ).

Formula (\ref{er2.12}) will be our fundamental representation for a tempered
ultradistribution. Sometimes use will be made of ``Dirac formula''
for exponential ultradistributions ( ref.\cite{tp6} ):
\begin{equation}
\label{er2.13}
\hat{F}(z)\equiv\frac {1} {2\pi i}\int\limits_{-\infty}^{\infty}
\frac {\hat{f}(t)} {t-z}\;dt\equiv
\frac {\cosh(\lambda z)} {2\pi i}\int\limits_{-\infty}^{\infty}
\frac {\hat{f}(t)} {(t-z)\cosh(\lambda t)}\;dt
\end{equation}
where the ``density'' $\hat{f}(t)$ is such that
\begin{equation}
\label{er2.14}
\oint\limits_{\Gamma} \hat{F}(z) \hat{\phi}(z)\;dz =
\int\limits_{-\infty}^{\infty} \hat{f}(t) \hat{\phi}(t)\;dt
\end{equation}
(\ref{er2.13}) should be used carefully.
While $\hat{F}(z)$ is analytic on $\Gamma$, the density $\hat{f}(t)$ is in
general singular, so that the r.h.s. of (\ref{er2.14}) should be interpreted
in the sense of distribution theory.

Another important property of the analytic representation is the fact
that on $\Gamma$, $\hat{F}(z)$ is bounded by an exponential and a power of $z$
( ref.\cite{tp6,tp8} ):
\begin{equation}
\label{er2.15}
|\hat{F}(z)|\leq C|z|^pe^{p|\Re(z)|}
\end{equation}
where $C$ and $p$ depend on $\hat{F}$.

The representation (\ref{er2.12}) implies that the addition of any entire function
$\hat{G}(z)\in{N}$ to $\hat{F}(z)$ does not alter the ultradistribution:
\[\oint\limits_{\Gamma}\{\hat{F}(z)+\hat{G}(z)\}\hat{\phi}(z)\;dz=
\oint\limits_{\Gamma} \hat{F}(z)\hat{\phi}(z)\;dz+\oint\limits_{\Gamma}
\hat{G}(z)\hat{\phi}(z)\;dz\]
But:
\[\oint\limits_{\Gamma} \hat{G}(z)\hat{\phi}(z)\;dz=0\]
as $\hat{G}(z)\hat{\phi}(z)$ is entire analytic
( and rapidly decreasing ),
\begin{equation}
\label{er2.16}
\therefore \;\;\;\;\oint\limits_{\Gamma} \{\hat{F}(z)+\hat{G}(z)\}\hat{\phi}(z)\;dz=
\oint\limits_{\Gamma} \hat{F}(z)\hat{\phi}(z)\;dz
\end{equation}

Another very important property of ${\large{B}}$ is that
${\large{B}}$ is reflexive under the Fourier transform:
\begin{equation}
\label{er2.17}
{\large{B}}={\cal F}_c\left\{{\large{B}}\right\}=
{\cal F}\left\{{\large{B}}\right\}
\end{equation}
where the complex Fourier transform $F(k)$ of $\hat{F}(z)\in{\large{B}}$
is given by:
\[F(k)=\Theta[\Im(k)]\int\limits_{{\Gamma}_+}\hat{F}(z)e^{ikz}\;dz-
\Theta[-\Im(k)]\int\limits_{{\Gamma}_{-}}\hat{F}(z)e^{ikz}\;dz=\]
\begin{equation}
\label{er2.18}
\Theta[\Im(k)]\int\limits_0^{\infty}\hat{f}(x)e^{ikx}\;dx-
\Theta[-\Im(k)]\int\limits_{-\infty}^0\hat{f}(x) e^{ikx}\;dx
\end{equation}
Here ${\Gamma}_+$ is the part of $\Gamma$ with $\Re(z)\geq 0$ and
${\Gamma}_{-}$ is the part of $\Gamma$ with $\Re(z)\leq 0$
Using (\ref{er2.18}) we can interpret Dirac's formula as:
\begin{equation}
\label{er2.19}
F(k)\equiv\frac {1} {2\pi i}\int\limits_{-\infty}^{\infty}
\frac {f(s)} {s-k}\; ds\equiv{\cal F}_c\left\{{\cal F}^{-1}\left\{f(s)\right\}\right\}
\end{equation}
The treatment for ultradistributions of exponential type defined on
${\boldsymbol{\mathbb{C}}}^n$ is similar to the case of one variable.
Thus
\begin{equation}
\label{er2.20}
{\Lambda}_j=\left\{z=(z_1, z_2,...,z_n)\in{\boldsymbol{\mathbb{C}}}^n :
|\Im(z_k)|\leq j\;\;\;1\leq k\leq n\right\}
\end{equation}
\begin{equation}
\label{er2.21}
||\hat{\phi} ||_j=\max_{k\leq j}\left\{\sup_{z\in{\Lambda}_j}\left[
e^{j\left[\sum\limits_{p=1}^n|\Re(z_p)|\right]}\left| D^{(k)}\hat{\phi}(z)\right|\right]\right\}
\end{equation}
where $D^{(k)}={\partial}^{(k_1)}{\partial}^{(k_2)}\cdot\cdot\cdot{\partial}^{(k_n)}\;\;\;\;
k=k_1+k_2+\cdot\cdot\cdot+k_n$

${{\large{B}}}^n$ is characterized as follows. Let
${{E}}^n_{\omega}$ be the space of all functions $\hat{F}(z)$ such that:

${\Large {\boldsymbol{I}}}^{'}$-
$\hat{F}(z)$ is analytic for $\{z\in \boldsymbol{{\mathbb{C}}^n} :
|Im(z_1)|>p, |Im(z_2)|>p,...,|Im(z_n)|>p\}$.

${\Large {\boldsymbol{II}}}^{'}$-
$\hat{F}(z)e^{-\left[p\sum\limits_{j=1}^n|\Re(z_j)|\right]}/z^p$
is bounded continuous  in
$\{z\in \boldsymbol{{\mathbb{C}}^n} :|Im(z_1)|\geqq p,|Im(z_2)|\geqq p,
...,|Im(z_n)|\geqq p\}$,
where $p=0,1,2,...$ depends on $\hat{F}(z)$.

Let ${{N}}^n$ be:
${{N}}^n=\left\{\hat{F}(z)\in{{E}}^n_{\omega} :\hat{F}(z)\;\right.$
is entire analytic at least in one of the variables $\left. z_j\;\;\;1\leq j\leq n\right\}$
Then ${{\large{B}}}^n$ is the quotient space:

${\Large {\boldsymbol{III}}}^{'}$-
${{\large{B}}}^n={{E}}^n_{\omega}/{{N}}^n$
We have now
\begin{equation}
\label{er2.22}
\hat{F}(\hat{\phi})=<\hat{F}(z), \hat{\phi}(z)>=\oint\limits_{\Gamma} \hat{F}(z) \hat{\phi}(z)\;
dz_1\;dz_2\cdot\cdot\cdot dz_n
\end{equation}
$\Gamma={\Gamma}_1\cup{\Gamma}_2\cup ...{\Gamma}_n$
where the path ${\Gamma}_j$ runs parallel to the real axis from
$-\infty$ to $\infty$ for $Im(z_j)>\zeta$, $\zeta>p$ and back from
$\infty$ to $-\infty$ for $Im(z_j)<-\zeta$, $-\zeta<-p$.
(Again $\Gamma$ surrounds all the singularities of $\hat{F}(z)$ ).
The n-dimensional Dirac's formula is
\begin{equation}
\label{er2.23}
\hat{F}(z)=\frac {1} {(2\pi i)^n}\int\limits_{-\infty}^{\infty}
\frac {\hat{f}(t)} {(t_1-z_1)(t_2-z_2)...(t_n-z_n)}\;dt_1\;dt_2\cdot\cdot\cdot dt_n
\end{equation}
where the ``density'' $\hat{f}(t)$ is such that
\begin{equation}
\label{er2.24}
\oint\limits_{\Gamma} \hat{F}(z)\hat{\phi}(z)\;dz_1\;dz_2\cdot\cdot\cdot dz_n =
\int\limits_{-\infty}^{\infty} f(t) \hat{\phi}(t)\;dt_1\;dt_2\cdot\cdot\cdot dt_n
\end{equation}
and the modulus of $\hat{F}(z)$ is bounded by
\begin{equation}
\label{er2.25}
|\hat{F}(z)|\leq C|z|^p e^{\left[p\sum\limits_{j=1}^n|\Re(z_j)|\right]}
\end{equation}
where $C$ and $p$ depend on $\hat{F}$.

\section{The Case N$\rightarrow\infty$}

\setcounter{equation}{0}

When the number of variables of the argument of the Ultradistribution of 
Exponential type tends to infinity we define:
\begin{equation}
\label{ep3.1}
d\mu(x)=\frac {e^{-x^2}} {\sqrt{\pi}}dx
\end{equation}
Let $\hat{\phi}(x_1,x_2,...,x_n)$ be such that:
\begin{equation}
\label{ep3.2}
\idotsint\limits_{-\infty}^{\;\;\infty}|\hat{\phi}(x_1,x_2,...,x_n)|^2 d{\mu}_1d{\mu}_2...
d{\mu}_n<\infty
\end{equation}
where
\begin{equation}
\label{ep3.3}
d{\mu}_i=\frac {e^{-x_i^2}} {\sqrt{\pi}}dx_i
\end{equation}
Then by definition
$\hat{\phi}(x_1,x_2,...,x_n)\in L_2({\mathbb{R}}^n,\mu)$
and 
\begin{equation}
\label{ep3.4}
L_2({\mathbb{R}}^{\infty},\mu)=
\bigcup\limits_{n=1}^{\infty}L_2({\mathbb{R}}^n,\mu)
\end{equation}
Let $\hat{\psi}$ be given by
\begin{equation}
\label{ep3.5}
\hat{\psi}(z_1,z_2,...,z_n)={\pi}^{n/4}\hat{\phi}(z_1,z_2,...,z_n)
e^{\frac {z_1^2+z_2^2+...+z_n^2} {2}}
\end{equation}
where $\hat{\phi}\in {{\large{Z}}}^n$(the corresponding 
n-dimensional of ${\large{Z}}$).\\
Then by definition $\hat{\psi}(z_1,z_2,...,z_n)\in{\large{G}}({\mathbb{C}}^n)$,
\begin{equation}
\label{ep3.6}
{\large{G}}({\mathbb{C}}^{\infty})=\bigcup\limits_{n=1}^{\infty}
{\large{G}}({\mathbb{C}}^n)
\end{equation}
and the dual ${\large{G}}^{'}({\mathbb{C}}^{\infty})$ given by
\begin{equation}
\label{ep3.7}
{\large{G}}^{'}({\mathbb{C}}^{\infty})=\bigcup\limits_{n=1}^{\infty}
{\large{G}}^{'}({\mathbb{C}}^n)
\end{equation}
is the space of Ultradistributions of Exponential type.\\
The analog to (\ref{er2.11}) in the infinite dimensional case is:
\begin{equation}
\label{ep3.8}
{\Large{W}}=\boldsymbol{(}{\large{G}}({\mathbb{C}}^{\infty}),
L_2({\mathbb{R}}^{\infty},\mu), 
{\large{G}}^{'}({\mathbb{C}}^{\infty})\boldsymbol{)}
\end{equation}
Let us now define:
\begin{equation}
\label{ep3.9}
{\cal F}:{\large{G}}({\mathbb{C}}^{\infty})\rightarrow
{\large{G}}({\mathbb{C}}^{\infty})
\end{equation}
via the Fourier transform:
\begin{equation}
\label{ep3.10}
{\cal F}:{\large{G}}({\mathbb{C}}^n)\rightarrow
{\large{G}}({\mathbb{C}}^n)
\end{equation}
given by:
\begin{equation}
\label{ep3.11}
{\cal F}\{\hat{\psi}\}(k)=
\int\limits_{-\infty}^{\infty}\hat{\psi}(z_1,z_2,...,z_n)
e^{ik\cdot z+\frac {k^2} {2}}d{\rho}_1d{\rho}_2...d{\rho}_n
\end{equation}
where
\begin{equation}
\label{ep3.12}
d\rho(z)=\frac {e^{-\frac {z^2} {2}}} {\sqrt{2\pi}}\;dz
\end{equation}
we conclude that
\begin{equation}
\label{ep3.13}
{\large{G}}^{'}({\mathbb{C}}^{\infty})=
{\cal F}_c\{{\large{G}}^{'}({\mathbb{C}}^{\infty})\}=
{\cal F}\{{\large{G}}^{'}({\mathbb{C}}^{\infty})\}
\end{equation}
where in the one-dimensional case
\begin{equation}
\label{ep3.14}
{\cal F}_c\{\hat{\psi}\}(k)=
\Theta[\Im(k)]\int\limits_{{\Gamma}_+}\hat{\psi}(z)e^{ikz+\frac {k^2} {2}}\;d\rho-
\Theta[-\Im(k)]\int\limits_{{\Gamma}_{-}}\hat{\psi}(z)e^{ikz+\frac {k^2} {2}}\;d\rho
\end{equation}

\section{The Constraints for a Bradyonic Bosonic String}

\setcounter{equation}{0}

The constraints for a bradyonic bosonic string have been deduced
in ref.\cite{tq1}. As a consequence  
we can describe  the bosonic string by a system composed of a
Lagrangian, one constraint and two initial conditions:
\begin{equation}
\label{ep4.1}
\begin{cases}
{\cal L}=|{\dot{X}}^2-{X^{'}}^2|\\
(\dot{X}+X^{'})^2=0\\
X_{\mu}(\tau,0)-X_{\mu}(\tau,\pi)=0
\end{cases}
\end{equation}
or equivalently
\begin{equation}
\label{ep4.2}
\begin{cases}
{\cal L}=|{\dot{X}}^2-{X^{'}}^2|\\
(\dot{X}-X^{'})^2=0\\
X_{\mu}(\tau,0)-X_{\mu}(\tau,\pi)=0
\end{cases}
\end{equation}

\section{A representation of  the states of the closed bosonic string}

\setcounter{equation}{0}

\subsection*{The case n finite}

From ref.\cite{tq1} we have
\begin{equation}
\label{ep5.1}
a=-z\;\;\;;\;\;\;a^+=\frac {d} {dz}
\end{equation}
Then
\begin{equation}
\label{ep5.2}
[a,a^+]=1
\end{equation}
Thus we have a representation for creation and annihilation operators
of the states of the string. The vacuum state annihilated 
by $z_{\mu}$ is the UET $\delta(z_{\mu})$,
and the orthonormalized states obtained by sucessive application of 
$\frac {d} {dz_{\mu}}$ to $\delta(z_{\mu})$ are:
\begin{equation}
\label{ep5.3}
F_n(z_{\mu})=\frac {{\delta}^{(n)}(z_{\mu})} {\sqrt{n!}}
\end{equation}

A general state of the string can be writen as:
\[\phi(x,\{z\})=[a_0(x)+a^{i_1}_{\mu_1}(x){\partial}^{\mu_1}_{i_1}+
a^{i_1 i_2}_{\mu_1\mu_2}(x){\partial}^{\mu_1}_{i_1}{\partial}^{\mu_2}_{i_2}
+...+...\]
\begin{equation}
\label{ep5.4}
+a^{i_1i_2...i_n}_{\mu_1\mu_2...\mu_n}(x){\partial}^{\mu_1}_{i_1}
{\partial}^{\mu_2}_{\i_2}...{\partial}^{\mu_n}_{i_n}+...+...]
\delta(\{z\})
\end{equation}
where $\{z\}$ denotes $(z_{1\mu},z_{2\mu},...,z_{n\mu},...,....)$, and 
$\phi$ is a UET of compact support in the set of variables $\{z\}$.
The functions
$a^{i_1i_2...i_n}_{\mu_1\mu_2...\mu_n}(x)$
are solutions of
\begin{equation}
\label{ep5.5}
\Box a^{i_1i_2...i_n}_{\mu_1\mu_2...\mu_n}(x)=0
\end{equation}

\subsection*{The case n$\rightarrow \infty$}

In this case
\begin{equation}
\label{ep5.6}
a=-z\;\;\;;\;\;\;a^+=-2z+\frac {d} {dz}
\end{equation}
we have
\begin{equation}
\label{ep5.7}
[a,a^+]=1
\end{equation}
The vacuum state annihilated by $a$ is $\delta(z)e^{z^2}$. The orthonormalized
states obtained by sucessive application of $a^+$ are:
\begin{equation}
\label{ep5.8}
{\hat{F}}_n(z)=2^{\frac {1} {4}}{\pi}^{\frac {1} {2}}
\frac {{\delta}^{(n)}(z)e^{z^2}} {\sqrt{n!}}
\end{equation}

\section{The World Sheet Supersymmetric String}

\setcounter{equation}{0}

We take as starting point the action given by:
\begin{equation}
\label{ep6.1}
S=-\frac {1} {2\pi}\int\limits_{-\infty}^{\infty}
\int\limits_0^{\pi}
|{\partial}_{\upsilon}X_{\mu}(\sigma){\partial}^{\upsilon}X^{\mu}(\sigma)
-i\overline{\psi}^{\mu}(\sigma){\rho}^{\upsilon}{\partial}_{\upsilon}
{\psi}_{\mu}(\sigma)|\;d^2\sigma    
\end{equation}
where
\begin{equation}
\label{ep6.2}
{\rho}^0=\left(
\begin{array}{ll}
0 & -i \\ 
i  & \;\;0
\end{array}
\right)
\;\;\;\;\;
{\rho}^1=\left(
\begin{array}{ll}
0 & \;\;i \\ 
i  & \;\;0
\end{array}
\right)
\;\;\;\;\;
\psi=\left(
\begin{array}{l}
{\psi}_1 \\ 
{\psi}_2
\end{array}
\right)
\end{equation}
Following a similar treatment to that of ref.\cite{tq1} we have the constraints
\begin{equation}
\label{ep6.3}
\begin{cases}
(\dot{X}+X^{'})^2=0\\
\dot{\psi}+{\psi}^{'}=0
\end{cases}
\end{equation}
or
\begin{equation}
\label{ep6.4}
\begin{cases}
(\dot{X}-X^{'})^2=0\\
\dot{\psi}-{\psi}^{'}=0
\end{cases}
\end{equation}
Thus, to describe the superstring we have:
\begin{equation}
\label{ep6.5}
\begin{cases}
{\cal L}=
|{\partial}_{\upsilon}X_{\mu}(\sigma){\partial}^{\upsilon}X^{\mu}(\sigma)
-i\overline{\psi}^{\mu}(\sigma){\rho}^{\upsilon}{\partial}_{\upsilon}
{\psi}_{\mu}(\sigma)|\\
(\dot{X}+X^{'})^2=0\\
\dot{\psi}+{\psi}^{'}=0\\
X_{\mu}(\tau,0)-X_{\mu}(\tau,\pi)=0\\
{\psi}_{\mu}(\tau,0)-{\psi}_{\mu}(\tau,\pi)=0
\end{cases}
\end{equation}
or
\begin{equation}
\label{ep6.6}
\begin{cases}
{\cal L}=|{\partial}_{\upsilon}X_{\mu}(\sigma){\partial}^{\upsilon}X^{\mu}(\sigma)
-i\overline{\psi}^{\mu}(\sigma){\rho}^{\upsilon}{\partial}_{\upsilon}
{\psi}_{\mu}(\sigma)|\\
(\dot{X}-X^{'})^2=0\\
\dot{\psi}-{\psi}^{'}=0\\
X_{\mu}(\tau,0)-X_{\mu}(\tau,\pi)=0\\
{\psi}_{\mu}(\tau,0)-{\psi}_{\mu}(\tau,\pi)=0
\end{cases}
\end{equation}
If we define:
\begin{equation}
\label{ep6.7}
{\cal L}_1={\partial}_{\alpha}X_{\mu}(\sigma){\partial}^{\alpha}X^{\mu}(\sigma)
-i\overline{\psi}^{\mu}(\sigma){\rho}^{\alpha}{\partial}_{\alpha}
{\psi}_{\mu}(\sigma)
\end{equation}
the Euler-Lagrange equations for the string are:
\begin{equation}
\label{ep6.8}
\frac {\partial} {\partial\tau}[Sgn({\cal L}_1){\dot{X}}^{\mu}]-
\frac {\partial} {\partial\sigma}[Sgn({\cal L}_1){X}^{'\mu}]=0
\end{equation}
\begin{equation}
\label{ep6.9}
Sgn({\cal L}_1){\rho}^{\upsilon}{\partial}_{\upsilon}{\psi}^{\mu}=0
\end{equation}
Equation (\ref{ep6.9}) implies that $Sgn({\cal L}_1)\neq 0$. The 
solution to (\ref{ep6.5}) is
\begin{equation}
\label{ep6.10}
\begin{cases}
{\psi}^{\mu}_1=\sum\limits_{n=-\infty}^{\infty}
c^{\mu}_{1n}e^{-2in(\tau-\sigma)}\\
{\psi}^{\mu}_2=0\\
X^{\mu}=x^{\mu}+l^2 p^{\mu}\tau+
\frac {il} {2}
\sum\limits_{n=-\infty\;;\;n\neq 0}^{\infty}
\frac {a^{\mu}_n} {n} e^{-2in(\tau-\sigma)}\\
p^2|\Phi>=0
\end{cases}
\end{equation}
and for (\ref{ep6.6}) is:
\begin{equation}
\label{ep6.11}
\begin{cases}
{\psi}_1^{\mu}=0\\
{\psi}^{\mu}_2=\sum\limits_{n=-\infty}^{\infty}
c^{\mu}_{2n}e^{-2in(\tau+\sigma)}\\
X^{\mu}=x^{\mu}+l^2 p^{\mu}\tau+
\frac {il} {2}
\sum\limits_{n=-\infty\;;\;n\neq 0}^{\infty}
\frac {a^{\mu}_n} {n} e^{-2in(\tau+\sigma)}\\
p^2|\Phi>=0
\end{cases}
\end{equation}
where $|\Phi>$ is the physical state of the string.

We will show that (\ref{ep6.1}) is a supersymmetric invariant. 
For this pourpose we use the equality for UET:
\[-\frac {1} {2\pi i}(z+y)[\ln(z+y)+\ln(-z-y)]=\]
\[-\frac {1} {2\pi i}z[\ln(z)+\ln(-z)]-
 \frac {1} {2\pi i}[\ln(z)+\ln(-z)]y+\]
\begin{equation}
\label{ep6.12}
2\sum\limits_{n=-\infty}^{\infty}{\delta}^{(n)}(z)
\frac {y^{2+n}} {(2+n)!}
\end{equation}
that on the real axis transforms into:
\begin{equation}
\label{ep6.13}
|x+y|=|x|+Sgn(x)\;y+
2\sum\limits_{n=-\infty}^{\infty}{\delta}^{(n)}(x)
\frac {y^{2+n}} {(2+n)!}
\end{equation}
As is known supersymmetry transformatios are given by
(see ref.\cite{tp13}):
\begin{equation}
\label{ep6.14}
\begin{cases}
\delta X^{\mu}=\overline{\epsilon}{\psi}^{\mu}\\
\delta{\psi}^{\mu}=-i{\rho}^{\upsilon}{\partial}_{\upsilon}
x^{\mu}\epsilon 
\end{cases}
\end{equation}
To show the invariance we use:
\begin{equation}
\label{ep6.15}
\begin{cases}
 Sgn({\cal L}_1)\neq 0\\
\dot{\psi}\pm{\psi}^{'}=0
\end{cases}
\end{equation}
The variation of ${\cal L}$ is:
\begin{equation}
\label{ep6.16}
\delta{\cal L}=2Sgn({\cal L}_1)({\dot{X}}^{\mu}\overline{\epsilon}
{\dot{\psi}}_{\mu}-X^{'\mu}{\overline{\epsilon}}{\psi}^{'}_{\mu})
\end{equation}
and as a consequence:
\begin{equation}
\label{ep6.17}
\delta S=-2\int\limits_{-\infty}^{\infty}\int\limits_0^{\pi}
Sgn({\cal L}_1)
({\ddot{X}}^{\mu}-X^{''\mu})\overline{\epsilon}{\psi}_{\mu}=0
\end{equation}

\section{A representation of  the states of the closed supersymmatric string}

\setcounter{equation}{0}

\subsection*{The case n finite}

As in ref.\cite{tq1}, for n finite we have:
\begin{equation}
\label{ep7.1}
\begin{cases}
a=-z\;\;\;;\;\;\;a^+=\frac {d} {dz}\\
c=\frac {d} {d\theta}\;\;\;;\;\;\;c^+=\theta\\
\end{cases}
\end{equation}
\begin{equation}
\label{ep7.2}
[a,a^+]=\{c,c^+\}=1
\end{equation}
where $-z$ and $d/dz$ are operators over CUET and $\theta$ 
is a Grassman variable with scalar product defined by:
\begin{equation}
\label{ep7.3}
<f,g>=\int f(\theta_1)e^{\theta_1{\theta}_2}g(\theta_2)\;d\theta_1\;
d{\theta}_2
\end{equation}
In a similar way to that for the bosonic string, a general state of the
supersymmetric string can be writen as: 
\[{\Phi}(x,\{z\},\{\theta\})=[c_0a_0(x)+
c(1,0)a^{i_1}_{{\mu}_1}(x){\partial}^{{\mu}_1}_{i_1}+\]
\[c(0,1)a^{j_1}_{{\alpha}_1}(x){\theta}^{{\alpha}_1}_{j_1}+
\cdot\cdot\cdot+\]
\[c(m,n)a_{{\mu}_1\cdot\cdot\cdot{\mu}_m
{\alpha}_1\cdot\cdot\cdot{\alpha}_n}^{i_1\cdot
\cdot\cdot i_m j_1\cdot\cdot\cdot j_n}(x)
{\partial}^{{\mu}_1}_{i_1}
\cdot\cdot\cdot{\partial}^{{\mu}_m}_{i_m}
{\theta}^{{\alpha}_1}_{j_1}\cdot\cdot\cdot{\theta}^{{\alpha}_n}_{j_n}+\]
\begin{equation}
\label{ep7.4}
+\cdot\cdot\cdot+\cdot\cdot\cdot]\delta(\{z\})
\end{equation}
where $c(m,n)$ are constants to be determined.\\
$\Phi$ satisfy:
\begin{equation}
\label{ep7.5}
\Box\Phi(x,\{z\},\{\theta\})=0
\end{equation}
\begin{equation}
\label{ep7.6}
\Box a_{{\mu}_1\cdot\cdot\cdot{\mu}_m
{\alpha}_1\cdot\cdot\cdot{\alpha}_n}^{i_1\cdot
\cdot\cdot i_m j_1\cdot\cdot\cdot j_n}(x)=0
\end{equation}

\subsection*{The case n$\rightarrow \infty$}

In this case:
\begin{equation}
\label{ep7.7}
\begin{cases}
a=-z\;\;\;;\;\;\;a^+=-2z+\frac {d} {dz}\\
c=\frac {d} {d\theta}\;\;\;;\;\;\;c^+=\theta
\end{cases}
\end{equation}
\begin{equation}
\label{ep7.8}
[a,a^+]=\{c,c^+\}=1
\end{equation}
and the expression for the physical state of the string is
similar to that for the finite case.

\section{The Field of the Supersymmetric String }

\setcounter{equation}{0}

According to (\ref{ep6.10}),(\ref{ep6.11}) and section 7 the equation for the string field
is given by
\begin{equation}
\label{ep8.1}
\Box\Phi(x,\{z\},\{\theta\})=0
\end{equation}
where $\{z\}$ denotes $(z_{1\mu},z_{2\mu},...,z_{n\mu},...,....)$, and 
$\Phi$ is a CUET in the set of variables $\{z\}$.
Any UET of compact support can be writed as a development of
$\delta(\{z\})$ and its derivatives. Thus we have:
\[{\Phi}(x,\{z\},\{\theta\})=[c_0A_0(x)+
c(1,0)A^{i_1}_{{\mu}_1}(x){\partial}^{{\mu}_1}_{i_1}+\]
\[c(0,1)A^{j_1}_{{\alpha}_1}(x){\theta}^{{\alpha}_1}_{j_1}+
\cdot\cdot\cdot+\]
\[c(m,n)A_{{\mu}_1\cdot\cdot\cdot{\mu}_m
{\alpha}_1\cdot\cdot\cdot{\alpha}_n}^{i_1\cdot
\cdot\cdot i_m j_1\cdot\cdot\cdot j_n}
(x){\partial}^{{\mu}_1}_{i_1}
\cdot\cdot\cdot{\partial}^{{\mu}_m}_{i_m}
{\theta}^{{\alpha}_1}_{j_1}\cdot\cdot\cdot{\theta}^{{\alpha}_n}_{j_n}+\]
\begin{equation}
\label{ep8.2}
+\cdot\cdot\cdot+\cdot\cdot\cdot]\delta(\{z\})
\end{equation}
where $C)m,n)$ are constants to be determined and  the quantum fields \\
$A_{{\mu}_1\cdot\cdot\cdot{\mu}_m
{\alpha}_1\cdot\cdot\cdot{\alpha}_n}^{i_1\cdot
\cdot\cdot i_m j_1\cdot\cdot\cdot j_n}(x) $
are solutions of
\begin{equation}
\label{ep8.3}
\Box A_{{\mu}_1\cdot\cdot\cdot{\mu}_m
{\alpha}_1\cdot\cdot\cdot{\alpha}_n}^{i_1\cdot
\cdot\cdot i_m j_1\cdot\cdot\cdot j_n}(x)=0
\end{equation}
The propagator of the string field can be exppresed in terms of the propagators 
of the component fields:
\[{\Delta}(x-x^{'},\{z\},\{z^{'}\},\{\theta\},\{{\theta}^{'}\})
=[c_0^2
{\Delta}(x-x^{'})+\cdot\cdot\cdot+\]
\[c^2(m,n){\Delta}_{{\mu}_1\cdot\cdot\cdot{\mu}_m
{\alpha}_1\cdot\cdot\cdot{\alpha}_n
{\nu}_1\cdot\cdot\cdot{\nu}_m{\beta}_1\cdot\cdot\cdot{\beta}_n
}^{i_1\cdot\cdot\cdot i_m
j_1\cdot\cdot\cdot j_n k_1\cdot\cdot\cdot k_m l_1\cdot\cdot\cdot l_n
}(x-x^{'})\]
\[{\partial}^{{\mu}_1}_{i_1}\cdot\cdot\cdot{\partial}^{{\mu}_m}_{i_m}
{\partial}^{'{\nu}_1}_{k_1}\cdot\cdot\cdot{\partial}^{'{\nu}_m}_{k_m}
{\theta}^{{\alpha}_1}_{j_1}\cdot\cdot\cdot{\theta}^{{\alpha}_n}_{j_n}
{\theta}^{'{\beta}_1}_{l_1}\cdot\cdot\cdot
{\theta}^{'{\beta}_n}_{l_n}+\]
\begin{equation}
\label{ep8.4}
+\cdot\cdot\cdot]\delta(\{z\},\{z^{'}\})
\end{equation}
Writing
\[A_{{\mu}_1\cdot\cdot\cdot{\mu}_m
{\alpha}_1\cdot\cdot\cdot{\alpha}_n
}^{i_1\cdot
\cdot\cdot i_m j_1\cdot\cdot\cdot j_n
}(x)=\int\limits_{-\infty}^{\infty}
(a_{{\mu}_1\cdot\cdot\cdot{\mu}_m
{\alpha}_1\cdot\cdot\cdot{\alpha}_n
}^{i_1\cdot
\cdot\cdot i_m j_1\cdot\cdot\cdot j_n
}(k)e^{-ik_{\mu}x^{\mu}}+\]
\begin{equation}
\label{ep8.5}
b_{{\mu}_1\cdot\cdot\cdot{\mu}_m
{\alpha}_1\cdot\cdot\cdot{\alpha}_n
}^{+i_1\cdot
\cdot\cdot i_m j_1\cdot\cdot\cdot j_n
}(k)e^{ik_{\mu}x^{\mu}})
\;d^{\nu-1}k
\end{equation}
We define the operators of annihilation and creation of a string as:
\[a(k,\{z\},\{\theta\})=
[c_0a_0(k)+c(1,0)a_{\mu_1}^{i_1}(k)
\partial_{i_1}^{\mu_1}+\]
\[c(0,1)a_{{\alpha}_1}^{j_1}(k){\theta}^{{\alpha}_1}_{j_1}+
\cdot\cdot\cdot+\]
\[c(m,n)a_{{\mu}_1\cdot\cdot\cdot{\mu}_m
{\alpha}_1\cdot\cdot\cdot{\alpha}_n
}^{i_1\cdot
\cdot\cdot i_m j_1\cdot\cdot\cdot j_n
}(k){\partial}^{{\mu}_1}_{i_1}
\cdot\cdot\cdot{\partial}^{{\mu}_m}_{i_m}
{\theta}^{{\alpha}_1}_{j_1}\cdot\cdot\cdot{\theta}^{{\alpha}_n}_{j_n}
+\]
\begin{equation}
\label{ep8.6}
+...+...]\delta(\{z\})
\end{equation}
\[a^+(k,\{z\},\{\theta\})=
[c_0a^+_0(k)+c(1,0)a_{\mu_1}^{+i_1}(k)
\partial_{i_1}^{\mu_1}+\]
\[c(0,1)a_{{\alpha}_1}^{+j_1}(k){\theta}^{{\alpha}_1}_{j_1}+
+\cdot\cdot\cdot+\]
\[c(m,n)a_{{\mu}_1\cdot\cdot\cdot{\mu}_m
{\alpha}_1\cdot\cdot\cdot{\alpha}_n
}^{+i_1\cdot
\cdot\cdot i_m j_1\cdot\cdot\cdot j_n
}(k){\partial}^{{\mu}_1}_{i_1}
\cdot\cdot\cdot{\partial}^{{\mu}_m}_{i_m}
{\theta}^{{\alpha}_1}_{j_1}\cdot\cdot\cdot{\theta}^{{\alpha}_n}_{j_n}
+\]
\begin{equation}
\label{ep8.7}
+...+...]\delta(\{z\})
\end{equation}
where the constans $c(m,n)$ are solution of:
\[c^{\ast}(m,n)
{\theta}^{{\alpha}_n}_{j_n}\cdot\cdot\cdot{\theta}^{{\alpha}_1}_{j_1}
a_{{\mu}_1\cdot\cdot\cdot{\mu}_m
{\alpha}_1\cdot\cdot\cdot{\alpha}_n
}^{+i_1\cdot
\cdot\cdot i_m j_1\cdot\cdot\cdot j_n
}(k)=\]
\begin{equation}
\label{ep8.8}
c(m,n)a_{{\mu}_1\cdot\cdot\cdot{\mu}_m
{\alpha}_1\cdot\cdot\cdot{\alpha}_n
}^{+i_1\cdot
\cdot\cdot i_m j_1\cdot\cdot\cdot j_n
}(k)
{\theta}^{{\alpha}_1}_{j_1}\cdot\cdot\cdot{\theta}^{{\alpha}_n}_{j_n}
\end{equation}
and define the creation and annihilation operators of the anti-string:
\[b^+(k,\{z\},\{\theta\})=
[c_0b^+_0(k)+c(1,0)b_{\mu_1}^{+i_1}(k)
\partial_{i_1}^{\mu_1}+\]
\[c(0,1)b_{{\alpha}_1}^{+j_1}(k){\theta}^{{\alpha}_1}_{j_1}
+\cdot\cdot\cdot+\]
\[c(m,n)b_{{\mu}_1\cdot\cdot\cdot{\mu}_m
{\alpha}_1\cdot\cdot\cdot{\alpha}_n
}^{+i_1\cdot
\cdot\cdot i_m j_1\cdot\cdot\cdot j_n
}(k){\partial}^{{\mu}_1}_{i_1}
\cdot\cdot\cdot{\partial}^{{\mu}_m}_{i_m}
{\theta}^{{\alpha}_1}_{j_1}\cdot\cdot\cdot{\theta}^{{\alpha}_n}_{j_n}
+\]
\begin{equation}
\label{ep8.9}
+...+...]\delta(\{z\})
\end{equation}
\[b(k,\{z\},\{\theta\})=
[c_0b_0(k)+c(1,0)b_{\mu_1}^{i_1}(k)
\partial_{i_1}^{\mu_1}+\]
\[c(0,1)b_{{\alpha}_1}^{j_1}(k){\theta}^{{\alpha}_1}_{j_1}
+\cdot\cdot\cdot+\]
\[c(m,n)b_{{\mu}_1\cdot\cdot\cdot{\mu}_m
{\alpha}_1\cdot\cdot\cdot{\alpha}_n
}^{i_1\cdot
\cdot\cdot i_m j_1\cdot\cdot\cdot j_n
}(k){\partial}^{{\mu}_1}_{i_1}
\cdot\cdot\cdot{\partial}^{{\mu}_m}_{i_m}
{\theta}^{{\alpha}_1}_{j_1}\cdot\cdot\cdot{\theta}^{{\alpha}_n}_{j_n}
+\]
\begin{equation}
\label{ep8.10}
+...+...]\delta(\{z\})
\end{equation}
As a consecuence we have
\[\Phi(x,\{z\},\{\theta\})=
\int\limits_{-\infty}^{\infty}
(a(x,\{z\},\{\theta\})
e^{-ik_{\mu}x^{\mu}}+\]
\begin{equation}
\label{ep8.11}
b^+(x,\{z\},\{\theta\})
e^{ik_{\mu}x^{\mu}})\;d^{\nu-1}x
\end{equation}
If we define
\begin{equation}
\label{ep8.12}
\begin{cases}
[\;\;\;,\;\;\;]_n=[\;\;\;,\;\;\;] ; \;\;\;n\;\;\; even\\
[\;\;\;,\;\;\;]_n=\{\;\;\;,\;\;\;\} ; \;\;\;n\;\;\; odd
\end{cases}
\end{equation}
with
\[[a_{{\mu}_1\cdot\cdot\cdot{\mu}_m
{\alpha}_1\cdot\cdot\cdot{\alpha}_n
}^{i_1\cdot
\cdot\cdot i_m j_1\cdot\cdot\cdot j_n
}(k),
a_{{\nu}_1\cdot\cdot\cdot{\nu}_m
{\beta}_1\cdot\cdot\cdot{\beta}_n
}^{+k_1\cdot
\cdot\cdot k_m l_1\cdot\cdot\cdot l_n
}(k^{'})]_n=\]
\begin{equation}
\label{ep8.13}
f_{{\mu}_1\cdot\cdot\cdot{\mu}_m
{\alpha}_1\cdot\cdot\cdot{\alpha}_n
{\nu}_1\cdot\cdot\cdot{\nu}_m
{\beta}_1\cdot\cdot\cdot{\beta}_n
}^{
i_1\cdot\cdot\cdot i_m
j_1\cdot\cdot\cdot j_n
k_1\cdot\cdot\cdot k_m
l_1\cdot\cdot\cdot l_n
}(k)\delta(k-k^{'})
\end{equation}
Then
\[\{a(k,\{z\},\{\theta\}),
a^+(k^{'},\{z{'}\},\{\theta^{'}\})\}=
[c_0^2f_0(k)\delta(k-k^{'})+\cdot\cdot\cdot+\]
\[c^2(m,n)f_{{\mu}_1\cdot\cdot\cdot{\mu}_m
{\alpha}_1\cdot\cdot\cdot{\alpha}_n
{\nu}_1\cdot\cdot\cdot{\nu}_m{\beta}_1\cdot\cdot\cdot{\beta}_n
}^{i_1\cdot\cdot\cdot i_m
j_1\cdot\cdot\cdot j_n k_1\cdot\cdot\cdot k_m
l_1\cdot\cdot\cdot l_n}(k)\delta(k-k^{'})\]
\[{\partial}^{{\mu}_1}_{i_1}\cdot\cdot\cdot{\partial}^{{\mu}_m}_{i_m}
{\partial}^{'{\nu}_1}_{k_1}\cdot\cdot\cdot{\partial}^{'{\nu}_m}_{k_m}
{\theta}^{{\alpha}_1}_{j_1}\cdot\cdot\cdot{\theta}^{{\alpha}_n}_{j_n}
{\theta}^{'{\beta}_1}_{l_1}\cdot\cdot\cdot
{\theta}^{'{\beta}_n}_{l_n}\]
\begin{equation}
\label{ep8.14}
+\cdot\cdot\cdot]
\delta(\{z\},\{z^{'}\})
\end{equation}
and for the anti-string
\[[b_{{\mu}_1\cdot\cdot\cdot{\mu}_m
{\alpha}_1\cdot\cdot\cdot{\alpha}_n
}^{i_1\cdot
\cdot\cdot i_m j_1\cdot\cdot\cdot j_n
}(k),
b_{{\nu}_1\cdot\cdot\cdot{\nu}_m
{\beta}_1\cdot\cdot\cdot{\beta}_n
}^{+k_1\cdot
\cdot\cdot k_m l_1\cdot\cdot\cdot l_n
}(k^{'})]_n=\]
\begin{equation}
\label{ep8.15}
g_{{\mu}_1\cdot\cdot\cdot{\mu}_m
{\alpha}_1\cdot\cdot\cdot{\alpha}_n
{\nu}_1\cdot\cdot\cdot{\nu}_m
{\beta}_1\cdot\cdot\cdot{\beta}_n
}^{
i_1\cdot\cdot\cdot i_m
j_1\cdot\cdot\cdot j_n
k_1\cdot\cdot\cdot k_m
l_1\cdot\cdot\cdot l_n
}(k)\delta(k-k^{'})
\end{equation}
Thus
\[\{b(k,\{z\},\{\theta\}),
b^+(k^{'},\{z{'}\},\{\theta^{'}\})\}=
c_0^2g_0(k)\delta(k-k^{'})+\cdot\cdot\cdot+\]
\[c^2(m,n)g_{{\mu}_1\cdot\cdot\cdot{\mu}_m
{\alpha}_1\cdot\cdot\cdot{\alpha}_n
{\nu}_1\cdot\cdot\cdot{\nu}_m{\beta}_1\cdot\cdot\cdot{\beta}_n
}^{i_1\cdot\cdot\cdot i_m
j_1\cdot\cdot\cdot j_n k_1\cdot\cdot\cdot k_m
l_1\cdot\cdot\cdot l_n}(k)\delta(k-k^{'})\]
\[{\partial}^{{\mu}_1}_{i_1}\cdot\cdot\cdot{\partial}^{{\mu}_m}_{i_m}
{\partial}^{'{\nu}_1}_{k_1}\cdot\cdot\cdot{\partial}^{'{\nu}_m}_{k_m}
{\theta}^{{\alpha}_1}_{j_1}\cdot\cdot\cdot{\theta}^{{\alpha}_n}_{j_n}
{\theta}^{'{\beta}_1}_{l_1}\cdot\cdot\cdot
{\theta}^{'{\beta}_n}_{l_n}\]
\begin{equation}
\label{ep8.16}
+\cdot\cdot\cdot]
\delta(\{z\},\{z^{'}\})
\end{equation}

\section{The Action for the Field of the Supersymmetric
String}

\setcounter{equation}{0}

\subsection*{The case n finite}

The action for the free supersymmetric closed string field is:
\[S_{free}=\iint
\oint\limits_{\{\Gamma_1\}}\oint\limits_{\{\Gamma_2\}}
 \int\limits_{-\infty}^{\infty}
{\partial}_{\mu}{\Phi}(x,\{z_1\},\{{\theta}_1\})e^{\{z_1\}{\cdot}\{z_2\}}
e^{\{{\theta}_1\}{\cdot}\{{\theta}_2\}}\]
\begin{equation}
\label{ep9.1}
{\partial}^{\mu}\Phi^+(x,\{z_2\},\{{\theta}_1\})
\;d^{\nu}x\;\{dz_1\}\{dz_2\}\{d{\theta}_1\}\{d\theta_2\}
\end{equation}
A possible interaction is given by:
\[S_{int}=\lambda\;\iiiint\oint\limits_{\{\Gamma_1\}}
\oint\limits_{\{\Gamma_2\}}
\oint\limits_{\{\Gamma_3\}}
\oint\limits_{\{\Gamma_4\}}
\int\limits_{-\infty}^{\infty}
{\Phi}^+(x,\{z_1\},\{{\theta}_1\})
e^{\{z_1\}{\cdot}\{z_2\}}
e^{\{{\theta}_1\}{\cdot}\{{\theta}_2\}}\]
\[\Phi(x,\{z_2\},\{{\theta}_2\})
e^{\{z_2\}{\cdot}\{z_3\}}
e^{\{{\theta}_2\}{\cdot}\{{\theta}_3\}}
{\Phi}^+(x,\{z_3\},\{{\theta}_3\}\})
e^{\{z_3\}{\cdot}\{z_4\}}
e^{\{{\theta}_3\}{\cdot}\{{\theta}_4\}}\]
\begin{equation}
\label{ep9.2}
\Phi(x,\{z_4\},\{{\theta}_4\})\;
d^{\nu}x\;\{dz_1\}\{dz_2\}\{dz_3\}\{dz_4\}
\{d{\theta}_1\}\{d\theta_2\}
\{d{\theta}_3\}\{d\theta_4\}
\end{equation}
Both, $S_{free}$ and $S_{int}$ are non-local as expected.

\subsection*{The case n$\rightarrow \infty$}

In this case:
\[S_{free}=\iint
\oint\limits_{\{\Gamma_1\}}\oint\limits_{\{\Gamma_2\}}
 \int\limits_{-\infty}^{\infty}
{\partial}_{\mu}{\Phi}(x,\{z_1\},\{{\theta}_1\})e^{\{z_1\}{\cdot}\{z_2\}}
e^{\{{\theta}_1\}{\cdot}\{{\theta}_2\}}\]
\begin{equation}
\label{ep9.3}
{\partial}^{\mu}\Phi^+(x,\{z_2\},\{{\theta}_1\})
\;d^{\nu}x\;\{d{\eta}_1\}\{d{\eta}_2\}\{d{\theta}_1\}\{d\theta_2\}
\end{equation}
where
\begin{equation}
\label{ep9.4}
d\eta(z)=\frac {e^{-z^2}} {\sqrt{2}\;\pi} dz
\end{equation}
and
\[S_{int}=\lambda\;\iiiint\oint\limits_{\{\Gamma_1\}}
\oint\limits_{\{\Gamma_2\}}
\oint\limits_{\{\Gamma_3\}}
\oint\limits_{\{\Gamma_4\}}
\int\limits_{-\infty}^{\infty}
{\Phi}^+(x,\{z_1\},\{{\theta}_1\})
e^{\{z_1\}{\cdot}\{z_2\}}
e^{\{{\theta}_1\}{\cdot}\{{\theta}_2\}}\]
\[\Phi(x,\{z_2\},\{{\theta}_2\})
e^{\{z_2\}{\cdot}\{z_3\}}
e^{\{{\theta}_2\}{\cdot}\{{\theta}_3\}}
{\Phi}^+(x,\{z_3\},\{{\theta}_3\}\})
e^{\{z_3\}{\cdot}\{z_4\}}
e^{\{{\theta}_3\}{\cdot}\{{\theta}_4\}}\]
\begin{equation}
\label{ep9.5}
\Phi(x,\{z_4\},\{{\theta}_4\})\;
d^{\nu}x\;\{d{\eta}_1\}\{d{\eta}_2\}\{d{\eta}_3\}\{d{\eta}_4\}
\{d{\theta}_1\}\{d\theta_2\}
\{d{\theta}_3\}\{d\theta_4\}
\end{equation}

\subsection*{Gauge Conditions}

The gauge conditions for the string field are:
\begin{equation}
\label{ep9.6}
\int\oint\limits_{\{\Gamma\}}z_{i_1}^{\mu_1}\cdot\cdot\cdot
z_{i_k}^{\mu_k} {\partial}_{\mu_k}\cdot\cdot\cdot z_{i_m}^{\mu_m}
{\partial}_{\theta\; j_1}^{{\alpha}_1}\cdot\cdot\cdot
{\partial}_{\theta\; j_l}^{{\alpha}_l}\cdot\cdot\cdot
{\partial}_{\theta\; j_n}^{{\alpha}_n}
\Phi(x,\{z\},\{\theta\})\;\{dz\}\{\theta d\theta\}=0
\end{equation}
${\partial}_{{\mu}_k}=\partial / \partial x^{\mu_k}\;;\;
{\partial}_{\theta\;j_l}^{\alpha_l }=\partial / {\partial} {{\theta}_{\alpha_l}^{j_l} }
\;;\;1\leq k\leq m\;;\;m\geq 1\;;\;1\leq l\leq n\;;\;n\geq 1$. \\
With these gauge conditions the number of the components fields 
of the superstring field is finite, and the temporal components of all
fields are eliminated.

Another gauge conditions that can be added to (\ref{ep9.6}) are
\begin{equation}
\label{ep9.7}
\int\oint\limits_{\{\Gamma\}}z_{i_1}^{\mu_1}\cdot\cdot\cdot
z_{i_k}^{\mu_k}\cdot\cdot\cdot z_{i_m}^{\mu_m}
{\partial}_{\theta\; j_1}^{{\alpha}_1}\cdot\cdot\cdot
{\partial}_{\theta\; j_l}^{{\alpha}_l}\cdot\cdot\cdot
{\partial}_{\theta\; j_n}^{{\alpha}_n}
\Phi(x,\{z\},\{\theta\})\;\{dz\}\{\theta d\theta\}=0
\end{equation}
$1\leq k\leq m\;;\;m\geq 1\;;\;1\leq l\leq n\;;\;n\geq 1$. \\
These additional gauge conditions permit us nullify other 
component fields according to experimental data.
It should be noted that gauge conditions (\ref{ep9.6}) and
(\ref{ep9.7})  does not modify the movement equations of 
superstring field.

The convolution
of two propagators of  the string field is:
\begin{equation}
\label{ep9.8}
\hat{{\Delta}}_{\alpha\beta}(k,\{z_1\},\{z_2\},\{\theta_1\},\{{\theta}_2\},)\ast 
\hat{{\Delta}}_{\alpha\beta}(k^{'},\{z_3\},\{z_4\},\{\theta^{'}_1\},\{{\theta}^{'}_2\},)
\end{equation}
where $\ast$ denotes the convolution
of Ultradistributions of Exponential Type  
on the $k$ variable only.
With the use of the result
\begin{equation}
\label{ep9.9}
\frac {1} {\rho}\ast\frac {1} {\rho}=-\pi^2\ln\rho
\end{equation}
$(\rho=k_1^2+k_2^2+\cdot\cdot\cdot+k_{\nu}^2)$ 
in euclidean space
and
\begin{equation}
\label{ep9.10}
\frac {1} {\rho\pm i0}\ast\frac {1} {\rho\pm i0}=
 \mp i\pi^2\ln(\rho\pm i0)
\end{equation}
($\rho=k_0^2-k_1^2-\cdot\cdot\cdot-k_{\nu-1}^2)$ 
in minkowskian space,

\section{Discussion}

We have decided to begin this paper, for the benefit
of the reader, with a summary of the main characteristics
of Ultradistributions of Exponential Type and their Fourier
transform.

We have shown that UET are appropriate to
describe 
in a consistent way string and string field theories.
By means of  a new Lagrangian for the closed superstring 
we have obtained the  movement 
equation for the field of the string and solve it with the use of CUET.
We show that this string field 
is a linear superposition of CUET.
We give a definiton of anti-superstring.
We evaluate the propagator for the string field,
and calculate the convolution of two of them, taking
into account that string field theory is a non-local theory
of  UET of an infinite number of complex variables.
For practical calculations and experimental results 
we have given expressions that
involve only a finite number of variables.

As a final remark we would like to point out that our formulae
for convolutions follow from  general definitions. They are not
regularized expresions

\end{document}